\documentclass{svjour3}

\usepackage[usenames,dvipsnames]{color}
\usepackage{ulem}
\usepackage[utf8]{inputenc}

\usepackage{graphicx}

\usepackage{epsfig}

\usepackage{color}

\usepackage[table]{xcolor}

\newcommand{\delete}[1]{\sout{\textcolor{red}{}}}
\newcommand{\add}[1]{\textcolor{black}{#1}}

\begin{document}
%

\title{Priority Levels Based Multi-hop Broadcasting Method for Vehicular Ad~hoc Networks}

\titlerunning{Priority Levels Based Multi-hop Broadcasting Method for VANETs}


 \author{
   Wahabou Abdou \and Beno\^it Darties \and Nader Mbarek}
 		
   \institute {
    Wahabou Abdou \and Beno\^it Darties \and Nader Mbarek \at LE2I - UMR CNRS 6306, University of Burgundy, 9 Avenue Alain Savary, 21~000~Dijon~France\\\email{\{wahabou.abdou, benoit.darties, nader.mbarek\}@u-bourgogne.fr}
   }
   
   \authorrunning{W. Abdou, B. Darties, N. Mbarek} 
   
   \maketitle

\begin{abstract}
This paper deals with broadcasting problem in Vehicular Ad~hoc Networks (VANETs). This communication mode is commonly used for sending safety messages and
traffic information. However designing an efficient broadcasting protocol is hard to achieve since it has to take into account some parameters related to the
network environment, for example the network density, in order to avoid causing radio interferences.  In this paper, we propose a novel Autonomic
Dissemination Method (ADM) which delivers messages in accordance with given priority and density levels. The proposed approach is based on two steps: an offline
optimization process and an adaptation to the network characteristics. \add{The first step uses a genetic algorithm to find solutions that fit the network context. The second one relies on the Autonomic Computing paradigm. ADM allows each vehicle to dynamically adapt its broadcasting strategy not only with respect to the network density, but also
in accordance to the priority level of the message to send. The experimental results show that ADM effectively uses the radio resources even when there are globally many
messages to send simultaneously. Moreover, ADM allows to increase the message delivery
ratio and to reduce the latency and radio interferences.}
\end{abstract}


\section{Introduction}
\label{sect:intro}
 
A Vehicular Ad~hoc Network (VANET) is a collection of vehicles
communicating through wireless connections: each vehicle acts
simultaneously as a node and as a wireless router, allowing multi-hop packet
forwarding. Indeed, each node has a limited coverage area that contains the
neighbours it can directly communicate with. This area can vary from one hundred
meters to a few kilometers (depending on the wireless technology on-board,
external radio interferences,\dots). Using vehicle-to-vehicle
communications allows sending packets over wide distance through multi-hop
relays.  VANET are mainly characterized by a dynamic network topology
and a heterogeneous node density due to road traffic conditions.

This paper deals with broadcasting techniques which are used for sending safety
messages, traffic information or comfort messages. When a packet is
broadcasted, it is received by all nodes within the sender's coverage area
(provided that no interference or radio channel trouble occurs). Every receiver
will decide to relay or not the packet depending on its own broadcasting
strategy. This hop-to-hop communication would lead to a full coverage of the
network. Performing an efficient multi-hop broadcast in VANETs is however a
difficult task. The protocols should take into account the specificities of the
radio channel, the high mobility of nodes and the network density. The decision
to relay the packets is taken in a distributed way, but each node's decision has
a real impact on the efficiency of the overall dissemination strategy: in
high-density networks, too many relay nodes would quickly in\-crease the number
of collisions, leading to a saturation of the bandwidth and a significant
increase of the latency. On the other hand, if not enough relay tasks are
performed in low-density networks the message may not be widely disseminated. 
Several approaches are proposed in the literature to overcome this problem. \add{Some methods from Mobile Ad~hoc Networks (MANETs) use the neighborhood knowledge~\cite{scalable_broadcast_algorithm}~\cite{MPR_Minet} to choose the best relay nodes. However, the high mobility that characterises VANETs makes their application in such a network very difficult. Some stochastic methods use waiting time to reduce the number of redundant packets~\cite{karthikeyan10}~\cite{AckPBSM}~\cite{POCA}~\cite{GCGCRN13}~\cite{HBHF13}. But if this waiting time is not well chosen, it may increase the message dissemination time. The approach we present in this paper is based on the Smart-flooding protocol~\cite{MobilWare} that uses a genetic algorithm optimization process to dynamically adapt dissemination strategies with respect to the network density.}

It
is also important to adapt the broadcasting strategy to the priority level of
the message. For instance, emergency messages such accident alerts, should be
delivered as fast as can be done in the source node's neighbourhood. \delete{Conversely,
it does not matter if weather information are broadcasted with a more important
latency, since there is no emergency.} \add{Conversely, it does not matter if some weather information with limited impact and less urgent tourist information are broadcasted with a more important
latency since there is no emergency.}

This paper investigates the problem of building an autonomous and robust
broadcasting protocol, which provides each node with the adequate strategy to
determine if an incoming message has to be forwarded or not depending on its
priority level and the network density. The goal is to make effective use of
radio resources when there are many messages to send simultaneously.
The paper is organized as follows: Section~\ref{sect:related} discusses a state
of the art of existing broadcasting methods. Thereafter,
Section~\ref{sect:defBroadcast} formulates the broadcasting problem in VANETs as
a multiobjective problem and presents an optimization methodology. The
proposed Autonomic Dissemination Method (ADM) is detailed in
Section~\ref{sect:autonomicRobustBroadcast} and its performances are assessed
in Section~\ref{sect:xp}.

\section{Related Work}
\label{sect:related}

   \subsection{Broadcasting Protocols}
\label{subsect:relatedbroadcast}
 
In the literature, ad~hoc broadcasting methods are classified into two
categories: deterministic and stochastic methods.

  \subsubsection{Deterministic Methods}
  
\add{Deterministic Methods} are those for which the broadcasting process and the behaviour of each node is
predictable. 

The simplest broadcasting method is the \textit{Simple flooding}.
Every packet is relayed exactly once by each node. Thus, in a network
consisting of $n$ nodes, $n$ copies of the packet will be sent. A drawback of
this method is that it may lead to many useless redundant packets. Another well-known deterministic methods subcategory is made up of
\textit{neighbour knowledge-based} protocols.  These methods are based on a
comparison of lists of neighbours : 1-hop neighbour list for
\textit{Distributed Vehicular Broadcast}~(DV-CAST)~\cite{dv_cast} or 2-hop
neighbour list for
\textit{Scalable Broadcast Algorithm}~(SBA)~\cite{scalable_broadcast_algorithm}.
These lists are included in the broadcast packets so that the
receiver ($r$) can compare the sender's list to its own list. This comparison
allows to determine the additional nodes that may receive the message if it
is forwarded by $r$. Among \textit{neighbour knowledge-based} methods, the
\textit{Multi-point relay}~(MPR)~\cite{MPR_Minet} \delete{approach is  the
most famous. The MPR approach} consists in selecting, for each
node, the smallest set of its 1-hop neighbours that will allows connection with
all its 2-hop neighbours. 
If \textit{neighbour knowledge-based} methods can be considered as fairly 
accurate, their main drawback is their non-applicability in networks with
very high mobility, since information about neighbours become inaccurate
very quickly.

  \subsubsection{Stochastic Methods}

\add{Stochastic Methods} statistically assess the gain that could be obtained if the
packets are relayed by a given node. They include \textit{probabilistic schemes}
which try to limit the number of relays by setting~up the probability for each
node to relay the packets. For a given network density, there exists $p_s$, a
threshold value of probability, that would allow all
nodes receive the packets, while reducing the number of unnecessary repetitions
and causing few collisions. Any other value $p > p_s$ would not lead to better
coverage. One challenge is to determine the correct value of $p_s$.

Smart-flooding~\cite{MobilWare} is a probabilistic protocol that assigns,\add{among others}, to each node \delete{ some of parameters, including} the probability of retransmission and the
number of repetitions of each message. 
This protocol assumes that in some VANET
scenarios, a vehicle may have no neighbour when it sends a packet. Therefore,
it
may be necessary to repeat the packet several times. The parameters introduced
by Smart-flooding are optimized using a genetic
algorithm. This protocol has the distinction of being robust with respect to the
density in the case of sending an emergency message.
Smart-flooding \delete{served as inspiration}\add{inspires} for the contribution we present in this paper.

\textit{Counter-based methods} rely on the assumption that the more a node
receives copies of the packet, the less likely it is useful to relay this
packet. Upon reception of the first copy, the node initializes a $C$ counter and
sets a timeout RAD (Random Access Delay). During the waiting period, $C$
is incremented upon reception of a copy of the packet. When RAD expires, the
packet is relayed if $C$ is less than $C_t$, a threshold value. Like
probabilistic methods, the
challenge is to find the right value of $C_t$. 
Karthikeyan~et~al.~\cite{karthikeyan10}
proposed a protocol that defines two categories of nodes according to their
number of neighbours, with respect to a given threshold. Each node
decides to relay each packet depending on its own category and the category of
the last hop of this packet.

\textit{Location-based methods} relay messages, depending on the potential
additional coverage area that will result from this retransmission. These
technique do not consider whether nodes exist within that additional area or
not. AckPBSM~\cite{AckPBSM} and POCA~\cite{POCA} use this approach and set lower
RAD to nodes that are far from the source (or last-hop relay). \add{In~\cite{GCGCRN13}, Garcia-Lozano et al. use a continuous expression to compute waiting time in order to reduce the number of collisions. Their method is used for
advertising services like gas station location. They benefit form the advantages of distance-based methods in order to effectively use the bandwidth. The
authors add some mechanisms to allow their approach to cope with not fully connected networks (for example sparse networks during non-rush hours). They
use a store-carry forward mechanism. Our proposed protocol (ADM) handle this issue by repeating (if necessary) some packets many times. We note that Garcia-Lozano et
al.'s solution is for an application that convey information that is not critical. In the case of sending emergency messages, this approach could be penalized
by the waiting time that could increase the latency.}

\add{Bi-Zone Broadcast (BZB)~\cite{HBHF13} combines probabilistic and distance-based schemes to reduce both the dissemination delay and the overuse of the wireless
channel. According to a given threshold ($D_{th}$) the authors divide the area into two zones: nodes that are close to the source/relay node use a random
Contention-Based Forwarding (CBF); nodes that are beyond $D_{th}$ use a distance-based CBF. This leads to ensure that the farthest node has a lower waiting
time before forwarding the message. In case there is no node beyond $D_{th}$, the use of random CBF avoids stopping the broadcast process. In addition, the
authors discriminate potential relays based on their capabilities: Road-Side Units (RSUs) and tall vehicles are favoured during the relay selection
since they might have better antennas and higher transmission power.}

  \subsection{Autonomic Computing}
\label{subsect:relatedautonomous}
   
   Traditionally, networks management is a manually 
controlled process. The creation of self-management systems 
with limited human interventions was the vision to bring autonomy 
within IT environment in order to cope with increasing complexity 
and excessive maintenance costs. Networks become a collection of 
interconnected self-governed entities where human intervention 
is limited to high-level directives. The first initiative dealing 
with this new paradigm is inspired by biological systems and 
in particular, the autonomic nervous system~\cite{nader2_new}.
Although the objectives 
list of the self-management concept was extended since 2001 \delete{(year of this new paradigm birth)}, the main objectives for 
autonomic systems are Self-configuring, Self-optimizing, 
Self-healing and Self-protecting. To achieve those 
objectives, autonomic systems have a detailed knowledge of 
their internal state as well as their environment using a 
continuous monitoring of eventual changes that could affect 
their components. Detecting changes induces the autonomic system 
to adjust its resources and the monitoring continues to determine 
if the new measures satisfy the desired performance. That is the 
closed control loop of self-management systems. This loop is implemented by autonomic managers, which control managed 
resources using sensors and effectors manageability 
interfaces~\cite{nader5_new}.
 
We propose in this paper an autonomic  robust  broadcasting  method taking into account the Autonomic Computing concepts to adapt broadcasting strategies
thanks to a knowledge base provided by autonomic mobile nodes (see Section~\ref{sect:autonomicRobustBroadcast})

\section{The Broadcasting Problem}
\label{sect:defBroadcast}

   \subsection{Broadcasting in VANETs: an Optimization Problem}
\label{subsect:defBroadcastopti}

The broadcasting problem in VANETs is known to be NP-hard.
Designing an efficient protocol requires to meet several objectives that can be
antagonistic: transmitting messages to the maximum of nodes while
avoiding the overuse of the radio channel; delivering packets as quickly as
possible, knowing that this speed may cause radio interferences. In a nutshell,
this is clearly a multi-objective optimization problem for which each solution
is a set of parameters that define a broadcasting strategy.
This strategy  may consist of the following parameters:

\begin{itemize}
 \item $P$: the probability to relay every packet. Upon reception of the first
copy of a broadcast packet, each node decides to relay it or not, depending
on $P$. 
 \item $Nr$: the number of repetitions of each packet. In low density
networks, when a node sends a packet, it is not unusual that it has no
neighbour in its coverage area that will receive the message and relay it.
Therefore, it may be necessary to repeat the packet several times.
 \item $Dr$: the delay between two successive repetitions. Applicable only if
$Nr > 1$. A very short delay could lead to many interferences, whereas a very
long delay may slow down the broadcasting process.
 \item $TTL$: the Time To Live or the maximum number of hops for each packet.
The $TTL$ is used to limit the packets' spread. It could be
replaced by any parameter dealing with geographical coordinates or
transmission time.
\end{itemize}

The performance of broadcasting strategies is evaluated using four
critera:

\begin{itemize}
 \item the average number of Collisions ($NC$).
 \item the propagation Time ($PT$). It is the time between the transmission of
a packet and the time it is received by all nodes within the area of interest.
 \item the total number of repetitions of each packet ($R$). 
 \item the \textit{full reception ratio} ($FR$). It refers to the guarantee 
that the packets will be received by all nodes (the reachability). 
\end{itemize}

$NC$ and $R$ enable to measure the radio channel usage: high values indicate
that the evaluated stra\-tegy is likely to interfere with
other communications in the network.

Determining the best broadcasting strategy can be seen as a
multiobjective optimization problem that aims to find the
solution $\overrightarrow{x}$ 
(or a solution set ~$\overrightarrow{X}~=~\{\overrightarrow{x_0},~...,
~\overrightarrow{x_n}\}$) such that:

$
\overrightarrow{x}=[P, Nr, Dr, TTL] \ s.t.
\left\{
\begin{array}{l l}
  \mbox{$NC(\overrightarrow{x})$} & is\ to\ be\ minimized\\          
  \mbox{$PT(\overrightarrow{x})$}& is\ to\ be\ minimized\\
  \mbox{$R(\overrightarrow{x})$}& is\ to\ be\ minimized\\
  \mbox{$FR(\overrightarrow{x})$}& is\ to\ be\ maximized\\ 
\end{array} 
\right .
$

The next section describes the methodology which is used.

   \subsection{Methodology}
   \label{subsect:defMethodology}

\begin{figure}
	\begin{center}
	  \includegraphics[width=0.7\textwidth]{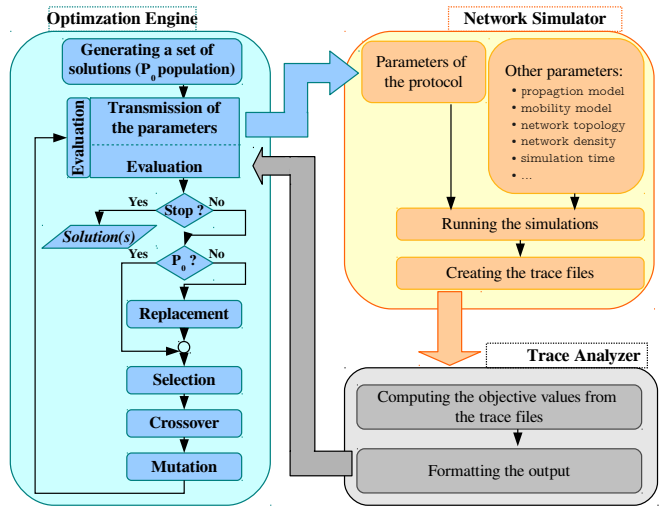}
	\end{center}
	\caption{The flowchart of HOPES}
	\label{fig:hopes}
\end{figure}

The $P$, $Nr$ $Dr$ and $TTL$ parameters are optimized using HOPES (Hybrid
Optimization Platform
using Evolutionary Algorithms and Simulations). Our platform combines an optimizer, a
network simulator and a trace analyzer. Figure~\ref{fig:hopes} illustrates the 
interaction of these three modules.

The Optimization Engine is used to effectively explore the search space  while
the Network Simulator assesses solution using models that are closed to reality.
We used aGAME\add{ }(adaptive Genetic Algorithm with Multiple parEto
sets)~\cite{GECCO_2012} as optimizer tool. aGAME generates a set of possible
solutions. Thereafter, theses solutions are transmitted to the network simulator
which integrates them with other parameters in order to better reproduce the
context of a real network. The trace files generated during the simulation are
then transmitted to the Trace Analyzer
module. The latter processes the trace files  in order to extract the values of
the objective functions ($NC$, $PT$, $R$, and $FR$) and presents the obtained
results according to format required by the genetic algorithm. Then, aGAME
uses these results to guide the exploration of the search space. The process is
repeated until a stop criterion is met (for instance the total number of
solutions to
evaluate).

The overall optimization process leads to a set of solutions, corresponding to
broadcasting strategies that fit a network with a given density level. This
process is repeated for several densities by changing the
corresponding parameter in the Network Simulator module. It is worth mentioning that this is an offline optimization process. The results
(that represents the ``best-suited'' broadcasting strategy for each density)
allow building a knowledge base that establishes 
a connection between density levels and broadcasting strategies. Each vehicle
can therefore choose the appropriate dissemination strategy, depending on the
density of the network in which it is located. The evaluation
of the density level is discussed in section~\ref{subsect:densityEvaluation}. 

\section{Density and Priority Levels Based Autonomic Dissemination Method}
\label{sect:autonomicRobustBroadcast}
 In this paper, we propose an extension of our Smart-flooding protocol thanks
to an autonomic robust broadcasting method called ADM (Autonomic Dissemination
Method). We adapt the broadcasting strategy used by the
Smart-flooding according to, not only the VANET's density level but also the
priority level of the message to disseminate. Indeed, ADM is based
on the closed control loop implemented by an autonomic manager within a mobile
node (vehicle).
 
  \subsection{Architecture}
    \label{subsect:autonomicRobustBroadcastArchitecture}
\begin{figure}[tbh!]
   \begin{center}
      \includegraphics[width=0.6\textwidth]{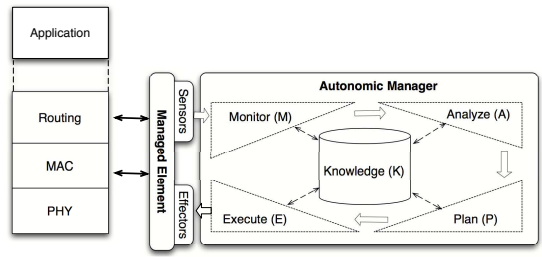}
   \end{center}
   \caption{Autonomic manager closed control loop}
   \label{fig:autonomic_loop}
\end{figure}

We adopt the self-management characteristics to improve robustness of 
Smart-flooding. Indeed, each node is considered as an
autonomic element thanks to an autonomic manager that enables broadcasting
decisions making according to environment changes in terms of density level (see
Section~\ref{subsect:densityEvaluation}) and takes into account message priority
level (see Section~\ref{subsect:autonomicRobustBroadcastPriority}). To achieve
these goals, the autonomic manager implements the MAPE-K closed control loop
(see Figure~\ref{fig:autonomic_loop}) and communicates with the mobile node
(called managed resource) using sensors and effectors manageability interfaces.

Each autonomic node within a VANET provides the Monitor function (M) of the
autonomic manager with network traffic information's thanks to the Sensors
manageability interface. In the context of the novel ADM protocol, the Monitor
determines if the received packet is a broadcasting one thanks to its
destination address. If so, the Monitor provides the Analyze function (A) with
this information to follow the control loop process. The Analyze function has to
determine, not only the priority level of the message according to the header
information's, but also the current density level of the node environment thanks
to the node local view table stored within the Knowledge base (K). After density
level evaluation  (detailed in Section~\ref{subsect:densityEvaluation}), the
Plan function (P) uses the density and priority values provided by the Analyze
function to retrieve the adequate broad\-cast strategy from the Knowledge base
thanks to the strategy table created by the offline op\-ti\-mi\-za\-tion phase
(see Section~\ref{subsect:defMethodology}). Then, the Plan function provides
the Execute function (E) with the broadcasting para\-meters (P, Nr, Dr and TTL)
in order to change the be\-havior of the mobile node managed resource by
executing the cor\-responding broadcast stra\-tegy  actions thanks to the
Effectors manageability interface.

  \subsection{Density Level Evaluation}
    \label{subsect:densityEvaluation}
      ADM evaluates the local density for each
autonomic node based on the number of active neighbours from which it received
\delete{the} packets. During communication, each node builds a view of its neighbourhood \add{based upon}\delete{. 
This view depends on} the neighbour list having transmitted or relayed packets.
Each autonomic node maintains a history in which it associates with each
received packet a list of nodes having sent or relayed it. Upon receipt of the
first copy of a packet, its identifier and the source/relay address are recorded
within the autonomic manager Knowledge base in a \delete{table called} local view \add{table}. When a
redundant copy is received, the identifier of the new relay is \delete{suffixed} \add{appended} to list
of addresses ($L$) corresponding to the packet. $L$ is stored in the local view
table. Each address is recorded only once for each packet. The current
number of neighbours ($\overline{N_i}$) for each $i$ autonomic node is equal to
the average number
of transmitters for all the packets stored in $L$ (see
Equation~\ref{eq:nb_neigh}).

\vspace{-0.12cm}
\begin{equation}
\overline{N_i} = \frac{\sum_{j=1}^{n}|L(j)|}{n}
 \label{eq:nb_neigh}
\end{equation}

where $n$ is the number of packets in the local view table and $|L(j)|$ 
the number of nodes that issued~/~relayed the $j^{th}$ packet in the table.

  \subsection{Priority Levels}
    \label{subsect:autonomicRobustBroadcastPriority}  

The messages' importance in VANETs leads to different priorities with particular
requirements. Introducing priority in broadcasted message has been investigated
in~\cite{Suthaputchakun}. 

\delete{We define three priority levels for broadcasted messages in VANETs 
and define for each level a broadcast policy to satisfy. These policies mainly
illustrate the adaptability of the protocol to the messages contents and can be
easily redefined or extended with other priority levels.}

\add{In this paper we focus on three priority levels for broadcasted messages in VANETs 
and we define for each level a broadcast policy to satisfy. The goal here is to highlight the capability of ADM to adapt to the messages contents. These priority levels could be
easily redefined or extended.}

\begin{itemize}
\item High-priority Level messages ($HL$), e.g. safety message or accident 
detection. They have to be delivered as quickly as possible since they may
require a prompt reaction from the driver. For these messages,
our protocol tries to minimize the required  propagation time, then to maximize
the full reception ratio. 
 \item Medium-priority Level messages ($ML$),  e.g.  traffic report. They
suppose less-critical information, where the driving reflexes are not part of
the equation and only attention is required. They should widely cover the
network while reducing the number of collisions.
\item Low-priority Level messages ($LL$), e.g. weather information, tourist
attraction or point of interest. They are optional messages whose delivery must
not alter the delivery of higher-priority messages. The use of the radio
resources  has to be optimized, though reducing the number of collisions as
well as the number of retransmissions, for an acceptable node coverage ratio.
\end{itemize}

\section{Experiments and Results}
\label{sect:xp}

\delete{In this paper, we focus on the ability of ADM to adapt its communication strategies to different message priority levels. We} \add{In this section, we} particularly look at the behaviour of \delete{this novel protocol} \add{ADM} when the
traffic load increases (many packets transmitted simultaneously). Due to a lack
of space, we will not discuss the robustness of ADM when varying the network
density.

The simulations were carried out using the ns2 network simulator (2.34~version), with Shadowing Pattern propagation
model~\cite{Shadowing_pattern}. It is a realistic and probabilistic propagation model which can produce statistical errors distributions, such as slow and fast fading, while being easy enough to be carried out on medium to large simulations.

Regarding the topology model, we considered a convoy of vehicles lined up on 10~kilometers. To illustrate different density levels, we varied the inter-vehicle distance. Table~\ref{tab:topo_params} shows the parameters of the topology used for different levels of density.

\begin{table}[h!]
\centering
\begin{tabular}{|p{2.65cm}|c|c|c|}
\hline

Density levels & Number of vehicles	& Inter-vehicle distance 	 & Number of neighbours	 \\ \hline
High (Urban)	&	400					& $25~m$						 & 26				\\ \hline
Medium (Suburban)	&	134					& $75~m$					& 10				\\ \hline
Low (Highway)	&	50					& $200~m$						 & 5				\\ \hline
Very low (Rural)	&	10					& $1000~m$					 & 1				\\ \hline

\end{tabular}
\caption{Topology parameters for different density levels}
\label{tab:topo_params}
\end{table}

The scenario of a very low-densty network illustrates inter-vehicle communications in rural areas. In these areas, vehicles rarely pass each other. Therefore they often have a few neighbours (if any). To represent such an environment, the \textit {Shadowing pattern} propagation model has been tuned so that each vehicle can communicate only over 20\% of the total simulation time. This corresponds roughly to a network with 10 vehciles lined up on 10 kilometers with an inter-vehicle distance of 1000 meters if the (classical) Shadowing propagation model is used.

Depending on the considered scenario (see Table~\ref{tab:topo_params}), each vehicle may have an average number of neighbours which varies from 1 to 26. 
Indeed, using WiFi, the broadcast packets are received over long
distances (up to several hundred meters, even
$1~km$~\cite{Schmidt10adaptingthe}). Since all messages are sent simultaneously and propagate in
a very short time, the mobility model is not relevant. Indeed, the propagation
time is less than 1~second. This situation prevents the network topology to
significantly change during communication.

  \subsection{Broadcasting Prameters Values for Each Priority Level}
  \label{subsect:xpparameters}
To determine the parameters of ADM for each priority level in various density \delete{levels} networks, we used the HOPES platform (see~Section~\ref{subsect:defMethodology}). Like in most
of multiobjective problems, the optimization process return\delete{ed}\add{s} as a
result several potential solutions which offer a compromise between the
different objective functions ($NC$, $PT$, $R$, $FR$). To refine the results,
we used a multiple-criteria decision-making approach based on preferences.

For sending high-priority messages, we select\delete{ed} the solution which
allow\delete{ed}\add{s} to deliver packets as quickly as possible while covering the
largest number of nodes in the network. For medium-priority messages, the first
criterion taken into account is the reachability ($FR$), then, among the
solutions that have a $FR$ value almost equal to 1 (the maximum), we select\delete{ed} the
one which causes the least collision. And finally, for the low-priority
messages, the goal is to send packets while slightly using the wireless channel.
The first and second criteria are respectively $NC$ \delete{et}\add{and} $R$. The broadcasting
parameters for the three priority levels and the objective functions values corresponding to various density levels are
presented in Tables~\ref{tab:parameters_uburban}~to~\ref{tab:parameters_rural}, respectively for high, medium, low and very low-density networks.
\add{For each scenario, we use one source node located at the end of the convoy of vehicles. Scenarios with multiple source nodes are discussed in Section~\ref{subsect:xpdiscussion}}.

\begin{table}[h!]
\centering
\begin{tabular}{|p{2.7cm}|c|c|c|c|c|c|c|c|c|}
\hline
&
\multicolumn{4}{c|}{{\bf Broadcasting parameters}}	& \cellcolor{black} &
\multicolumn{4}{c|}{{\bf Performance Results}}	\\ \hline
{\bf Message Classes} & $P$ & $Nr$ 
& $Dr$
& $TTL$ & \cellcolor{black} & $NC$	& $PT$	& $R$	& $FR$
\\ \hline
High-Level ($HL$) & 0.329	& 1
&
\cellcolor{lightgray}	& 32	 & \cellcolor{black} & 497	& 0.051 
& 131	& 99.6\%  	 \\
\hline
Medium-Level ($ML$) & 0.258	& 2	 & 1.721
& 15	 & \cellcolor{black} & 347	& 0.1063
& 207	& 100\%	 \\
\hline
Low-Level ($LL$)	 & 0.188	& 1
&
\cellcolor{lightgray}	& 39 & \cellcolor{black} & 190	& 0.048
& 75	& 86.8\% \\ \hline
\end{tabular}
\caption{ADM Parameters and Performance Results for a High-density Network (the Urban
Scenario)}
     	\label{tab:parameters_uburban}
\end{table}

In high-density networks, the probability to relay the packets \delete{are} \add{is} low (see
Table~\ref{tab:parameters_uburban}). When $Nr$ is equal to 1, the $Dr$ cell
(the delay between successive repetitions) has been darkened since this
parameter  is only applicable when $Nr > 1$. For high-priority messages (in the
high density network), relaying each packet only once, with a
probability of about 0.3 allows rapid dissemination of the message. However,
this probability value generates a large number of
collisions. This drawback is mended for medium-priority level messages. To
reduce the
number of collisions and increase the reachability ($FR$), we selected a
solution with a lower probability and a number of repetitions equal to 2.
Moreover, as the repetitions are not made in burst the risk of interference is
reduced.

For low-priority level messages, it is worth noting that the results only
concern the
packets that have been received by all vehicles. In other words, 86.8\% of
packets that are received spread quickly (due to low competition in the access
to the radio channel), but 13.2\% of them are not completely delivered.

Following the same reasoning, we obtain the broadcasting parameters for suburban
and highway scenarios
(Tables~\ref{tab:parameters_suburban}~and~\ref{tab:parameters_highway}
 respectively).

\begin{table}[h!]
\centering
\begin{tabular}{|p{2.7cm}|c|c|c|c|c|c|c|c|c|}
\hline
&
\multicolumn{4}{c|}{{\bf Broadcasting parameters}}	& \cellcolor{black} &
\multicolumn{4}{c|}{{\bf Performance Results}}	\\ \hline
{\bf Message Classes} & $P$ & $Nr$ 
& $Dr$
& $TTL$ & \cellcolor{black} & $NC$	& $PT$	& $R$	& $FR$
\\ \hline
High-Level  ($HL$) & 0.776	& 1	 &
\cellcolor{lightgray}	& 26	 & \cellcolor{black} & 166	& 0.044 
& 104	& 100\%  	 \\
\hline
Medium-Level ($ML$)	& 0.519	& 2	 & 0.951
& 16	 & \cellcolor{black} & 93	& 0.121
& 139	& 100\%	 \\ \hline
Low-Level ($LL$)	 & 0.291	& 2	 & 0.276
& 27 & \cellcolor{black} & 35
& 0.209	 & 82	& 75.8\%
\\ \hline
\end{tabular}
\caption{ADM Parameters and Performance Results for a Medium-density Network (the Suburban
Scenario)}
     	\label{tab:parameters_suburban}
\end{table}

\begin{table}[h!]
\begin{tabular}{|p{2.7cm}|c|c|c|c|c|c|c|c|c|}
\hline
&
\multicolumn{4}{c|}{{\bf Broadcasting parameters}}	& \cellcolor{black} &
\multicolumn{4}{c|}{{\bf Performance Results}}	\\ \hline
{\bf Message Classes}	 & $P$ & $Nr$ 
& $Dr$
& $TTL$ & \cellcolor{black} & $NC$	& $PT$	& $R$	& $FR$
\\ \hline
High-Level ($HL$)  	 & 0.999	& 4	 & 1.147
& 40	 & \cellcolor{black} & 31	& 0.092 
& 199	& 100\%  	 \\
\hline
Medium-Level  ($ML$)	 & 0.916	& 2	 & 0.729
& 28	 & \cellcolor{black} & 24	& 0.124
& 90	& 100\%	 \\ \hline
Low-Level  ($LL$)	 & 0.649	& 2	 & 1.933
& 34 & \cellcolor{black} & 10	& 1.414
& 66	& 82.8\% \\ \hline
\end{tabular}
\caption{ADM Parameters and Performance Results for a Low-density Network (the Highway 
Scenario)}
     	\label{tab:parameters_highway}
\end{table}

For the scenario of the rural area, the low density level of the network implies
the need to retransmit each packet many times (see
Table~\ref{tab:parameters_rural}). Indeed, in this scenario, VANETs behave like
delay tolerant networks (DTNs)~\cite{DTN_VANET}. In such a context, since the
radio channel is rarely used, even if ADM is able to differentiate broadcasting
strategies according to the class of a message, in practice these classes
scarcely
impact the communication process. The main constraints that must be met are:
having a probability $P$ close to 1 and a high number of repetition $Nr$.

\begin{table}[h!]
\begin{tabular}{|p{2.7cm}|c|c|c|c|c|c|c|c|c|}
\hline
&
\multicolumn{4}{c|}{{\bf Broadcasting parameters}}	& \cellcolor{black} &
\multicolumn{4}{c|}{{\bf Performance Results}}	\\ \hline
{\bf Message Classes} & $P$ & $Nr$ 
& $Dr$
& $TTL$ & \cellcolor{black} & $NC$	& $PT$	& $R$	& $FR$
\\ \hline
High-Level ($HL$) & 0.833	& 28	 & 0.233
& 28	 & \cellcolor{black} & 58	& 13.09 
& 1167	& 99.8\%  	 \\
\hline
Medium-Level ($ML$) & 0.896	& 25	 & 1.468
& 34	 & \cellcolor{black} & 16	& 28.295
& 1124	& 100\%	 \\ \hline
Low-Level ($LL$) & 0.902	& 8	 & 1.622
& 19 & \cellcolor{black} & 4	 & 30.957
& 362	 & 92.6\%
\\ \hline
\end{tabular}
\caption{ADM Parameters and Performance Results for a Very Low-density Network (the Rural Area Scenario)}
     	\label{tab:parameters_rural}
\end{table}

  \subsection{Performance evaluation}
     \label{subsect:xpdiscussion}

To assess the performance of ADM with respect to Simple flooding and Smart-flooding~\cite{MobilWare} methods, we considered the suburban scenario. Similar results are obtained for the other scenarios, but they are not presented in this article due to a lack of space.
 
Let us recall that the design of ADM had three major goals. ($g1$) swiftness: delivering of
high priority messages as soon as possible; ($g2$) network
coverage: reaching the maximum nodes for medium-priority messages; ($g3$)
effective use of radio channel for low priority messages. These objectives
must be met even if the traffic load increases (for instance when messages are
sent simultaneously). To better measure the achievement of these goals, we also
report on the results' figures, the performance of \textit{Simple flooding} and
\textit{Smart-flooding} protocols, for reference.

We vary the number of source nodes from 3 to 30. With 30 source nodes in a
convoy of vehicles over $10~km$, a message is issued approximately every
$330~meters$. Taking into account the communication range (for broadcast
messages), each node may have within its coverage area 4 or 5 neighbours which
simultaneously issue a message. At the second and third hop, the number of
simultaneously issued messages, within  each node's coverage area, greatly
increases. This may tend to quickly congest the radio channel.

\begin{figure}[b] 
 \centering
    \epsfig{figure =
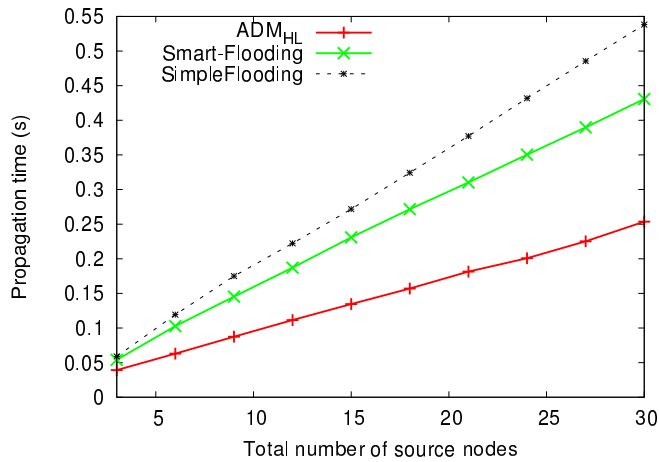,width=0.7\textwidth}
     \caption{Propagation time \label{fig:propagTime}}
 \end{figure}

Regarding the propagation time, ADM aims to deliver high-priority
messages (denoted ADM$_{HL}$) as fast as can be done, whatever the number of
source nodes. Figure~\ref{fig:propagTime} shows that the performances of ADM 
meet the first goal ($g1$). Compared to \textit{Smart-flooding}
and \textit{Simple-flooding}, ADM is less sensitive to the number of sources
than the other two protocols. Ultimately, even with 30 source nodes, the average
delay of priority messages is less than $250~ms$ (for a $10~km$ line), which is
acceptable. It is worth recall that the driver reaction time to traffic warning
signals can be in order of $700~ms$~\cite{robust_bcast}.

 \begin{figure}[!tb]
    \centering\epsfig{figure =
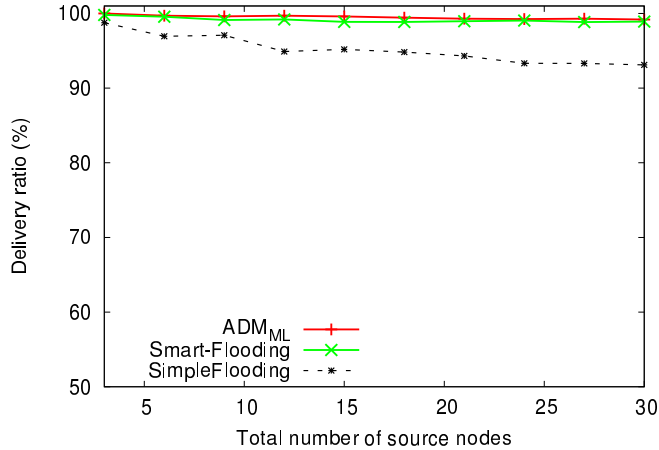,width=0.7\textwidth}
     \caption{Delivery ratio \label{fig:delivery}}
 \end{figure}
 
The goal $g2$ is assessed in Figure~\ref{fig:delivery} which shows the delivery
ratio. A packet is considered as ``delivered'' if it is received by all nodes.
Medium-priority packets (ADM$_{ML}$) are always received by all nodes. 
$ADM_{ML}$ ensures this result because it slightly increases the probability of
retransmission (see Table~\ref{tab:parameters_suburban}). 
It should be noted that if the value had been greatly increased, the
performance of ADM would be degraded and would get close \textit{Simple
flooding}'s results.

\begin{figure}[tb]    
    \centering\epsfig{figure
=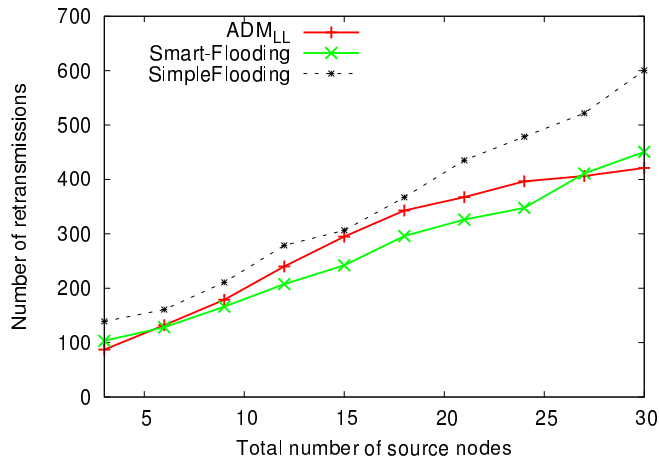,width=0.7\textwidth}
     \caption{Number of retransmissions \label{fig:retrans}}
 \end{figure}

\begin{figure}[hpt]    
    \centering\epsfig{figure =
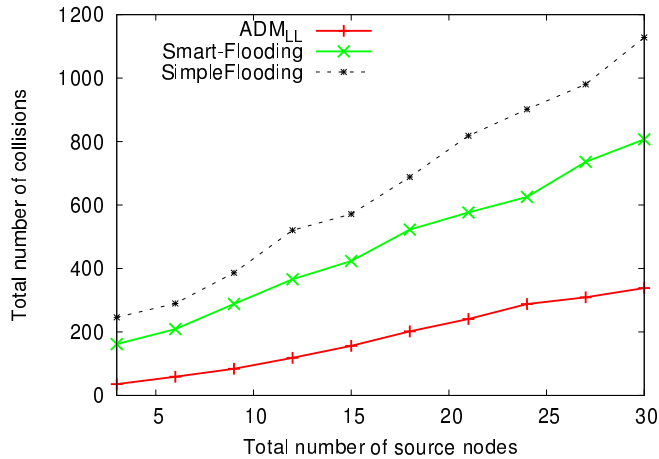,width=0.7\textwidth}
     \caption{Number of collisions \label{fig:col}}
 \end{figure}

And finally, Figures~\ref{fig:retrans}~and~\ref{fig:col} show that our third
goal ($g3$) is met: the low-priority messages use little radio channel by
limiting the total number of repetitions for each packet
(Figure~\ref{fig:retrans}). In addition, the fact that two potential successive
repetitions of the same packet are spaced out ($Dr$ value in
Table~\ref{tab:parameters_suburban}) reduces the number of collisions
(Figure~\ref{fig:col}).
 
\begin{figure}[tbh!]
   \begin{center}
       \includegraphics[width=0.7\textwidth]{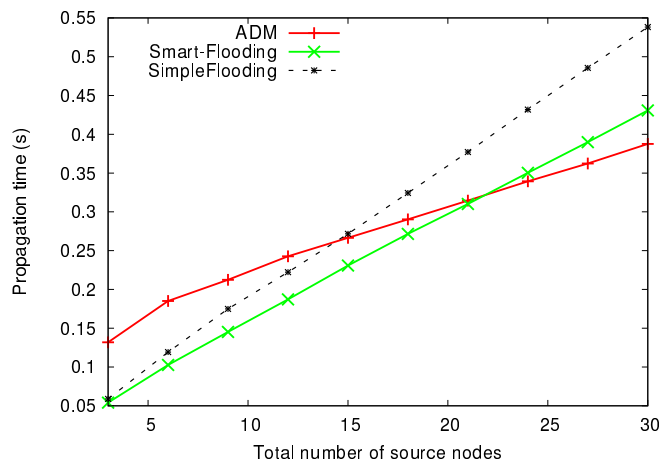}
   \end{center}
   \caption{Propagation time: ADM all priorities levels combined}
   \label{fig:cmp_propagTime}
\end{figure}

\begin{figure}[tbh!]
   \begin{center}
       \includegraphics[width=0.7\textwidth]{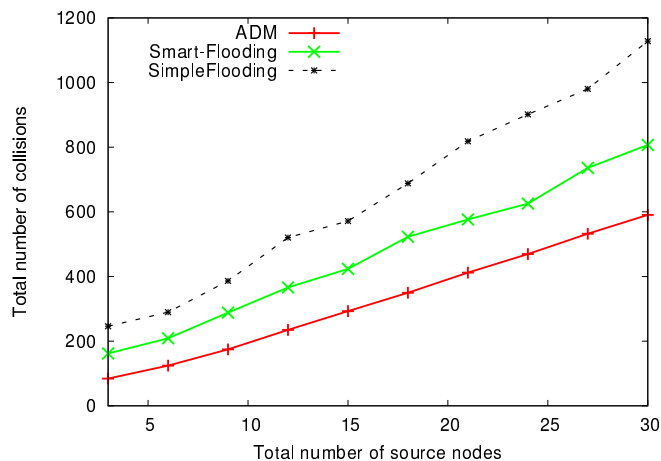}
   \end{center}
   \caption{Number of collisions: ADM all priorities levels combined}
   \label{fig:cmp_col}
\end{figure}

For a fair comparison, we also compare ADM to \textit{Smart-flooding} and \textit{Simple flooding} regarless the message priority levels. This means the results of ADM in these simulations (see Figures~\ref{fig:cmp_propagTime}~and~\ref{fig:cmp_col}) correspond to the average for all the messages (all priorities levels combined).

Figure~\ref{fig:cmp_propagTime} shows that if we consider the average delay of all packets, the overall performance of ADM is very interesting when the load of
the radio channel is high (many source nodes). This is explained by the key principle of ADM: when a packet's priority level is not high, ADM favors
broadcasting strategies that avoid overusing the radio channel. Thus when there \add{are} many concurrent access to the radio channel (many source nodes),
\textit{Smart-flooding} and \textit{Simple flooding} cause many collisions. Therefore the packets' propagation is slow\add{ed} down. This is corroborated  by
Figure~\ref{fig:cmp_col} which shows that the number of collisions when using ADM is lower than the other two flooding methods.

\section{Conclusion}    
      \label{sect:conclusion}

\add{This paper introduced an autonomic dissemination protocol (named ADM) that adapts broadcasting strategies to both network density and message priority level. This protocol relies on a genetic algorithm to optimize broadcasting strategies and Autnomic Computing concepts to adapt communication parameters according to the network context.} 
The simulations results reveal the scalability of ADM on a short-term period
when the number of simultaneous transmissions significantly increases. These
results also show that ADM outperforms two other broadcasting methods: the
Smart-flooding protocol and the simple flooding method. As an ongoing work, we are currently evaluating ADM performances when the network density level changes over the time. The aim is to establish the reliablity of ADM over a long communication period for various mobility models.

%



%

\bibliographystyle{spmpsci}


\end{document}